# In Europe

Jeroen van Dongen[*]

As the History of Science Society, which is based in America, holds its annual meeting in Utrecht, one of the key academic centers on the European continent, one may surmise that the field has returned home. Yet, this hardly reflects how today's world of scholarship is constituted: in the historiography of science, "provincializing Europe" has become an important theme, while the field itself, as is the case across the world of academia, is centered around a predominantly American literature. At the same time, ever since historians of science have emancipated themselves from the sciences a long time ago, they often have appeared, in the public eye, to question rather than to seek to bolster the authority of the sciences. How has this situation come about, and what does it tell us about the world we live in today? What insight is sought and what public benefit is gained by the historical study of science? As we try to answer these questions, we will follow a number of key mid-twentieth century historians—Eduard Dijksterhuis, Thomas Kuhn and Martin Klein—in their Atlantic crossings. Their answers to debates on the constitution of the early modern scientific revolution or the novelty of the work of Max Planck will illustrate how notions of "center" and "periphery" have shifted—and what that may tell us about being "in Europe" today.

It is a great honor for me to be invited by the History of Science Society (HSS) and the organizers of this year's HSS conference to present the Elisabeth Paris Lecture. Today I wish to address the unusual circumstance that the premier annual gathering of historians of science, the annual History of Science Society meeting, is taking place in the Netherlands; in fact, in Utrecht, to be precise. This is unusual, for the History of Science Society is based in the United States, and traditionally, for many years, has held its meetings exclusively in North America. So, why is it taking place in Utrecht? In the Netherlands? In Europe? Of course, there are particularities that shaped the decision to come here this week, but does that decision perhaps also reflect a larger state of affairs? May it tell us something about the world of scholarship today? Further, by raising that question, can we perhaps also learn something about the public role of historiography in that world? And finally, may the presence of the HSS here reveal us something about what it means to be "in Europe," as historians, today?

      Yet, is it really something worth noting that the History of Science Society is hosting its annual conference in Utrecht? Surely this is the major event in the field, and a substantial date on the calendar of the world of scholarship at large. Indeed, we have seen 1,000 participants

---

from all over the world congregate here this week, among them many from the United States. But, still, is the presence of the HSS meeting in Utrecht a circumstance that merits further thought? After all, in Utrecht we find one of the most prestigious continental European universities, which has enjoyed a long history of excellent scholarship—does this audience need reminding that it once terminated René Descartes' teaching stint early, because it felt it could do just fine without his services?

Still, is Utrecht, here and now, at the center of the global field of historiography of science? Can that center comfortably be located in Europe, today? Or is the fact that the HSS came to Utrecht perhaps an expression of a different circumstance: of a location on the periphery, despite Utrecht's intrinsic qualities? Again, these qualities are abundantly in evidence: Utrecht just saw off *Isis*, the premier journal of the field that has been very ably edited here; just as this university hosts a state-of-the-art master degree program and research center in history and philosophy of science, the Descartes Center, that is so graciously welcoming us here. If ever there was a center for the history of science, why not Utrecht in 2019?

Does it actually matter, to determine what is in the "center" and, its counterpoint, the periphery? Historians of science think it does: much scholarship has been dedicated to studies that aspire to "decenter" a location, epoch, or perspective. Such studies reveal how our sense of center, how our conditioning in the here and now, shapes, or perhaps biases our view of past and present. For example, two-and-a-half decades ago an important article by Andrew Cunningham and Perry Williams argued that one needed to "Decenter the Big Picture" in the history of science. What they meant concretely was that the beginning of a notion of "science" should not be placed in a "Scientific Revolution," which would have taken place circa the seventeenth century around figures like Nicolaus Copernicus, Galileo Galilei, and Isaac Newton, but that its beginning ought to be identified much later—since only sometime during the nineteenth century could one speak of science as the constellation of practices and values that we more or less understand it to be today. Placing its origin at the earlier point, in the seventeenth century, was in fact a side effect of the *invention of science* in the nineteenth century; it was the accompanying invention of a creation story, so to speak. They argued that doing so distorted and obscured from view what scholars in the seventeenth century were actually doing; one chose to ignore that Newton was a dedicated alchemist, too, for example.[1] In any case, Cunningham and Williams's invocation to "decenter" has a broader reach, beyond the intricacies of the early modern scientific revolution. It is this broader reach that I wish to address today.

Cunningham and Perry pointed out, as others did before and after them,[2] that the notion of science as a universal and timeless way to know the world has been conditioned by the here and now; the presumption of science's universality reflects how we believe today that one should properly know the world. Furthermore, they argued that the historiographical emphasis on Europe has been a consequence of that conditioning. "Decentering," that is, understanding how our historiography has been shaped by our own cultural conditioning and then seeking to look beyond that conditioning, is then necessary. This is necessary to acquire a full perspective on the historical acquisition of knowledge, and the place of "science" in that history. In this effort, one may seek to "provincialize Europe"; that is, one may seek to see how historically the European notion of rationality itself had come to be seen as universal and self-evident, also beyond its geographical location.[3]

Histories of global knowledge—and I consider myself part of that movement too—aspire to arrive at an understanding of knowledge creating cultures that cut across boundaries of continent, social stratification, or discipline. Doing away with such boundaries offers rich new insights: it shows connections and conditions that would otherwise be overlooked or taken for granted. It has revealed, for example, that the humanities and sciences have a much more closely interconnected history than a science-centered historiography might have shown[4]—and it has famously made clear how science and medicine, for instance through projects of sanitation or development, have been extensions and expressions of global power relations, most notably in colonial and postcolonial contexts.[5]

A focus on global power relations may yet tell us something about the position of Utrecht. As students of the reconstruction of postwar European scholarship are well aware, the United States actively promoted the regeneration of European science, using such instruments as Marshall aid or Fulbright scholarships. This exercise of soft power was in part intended to create a strong Western "arsenal of knowledge" to draw upon in the Cold War context.[6] It also aimed to further the acceptance of American leadership in Europe through cultural exchange.[7] Its effect has been the integration of European science in the American knowledge infrastructure. In the case of the Netherlands, that co-construction was very much sought: scholarly integration was seen as a vital way to position oneself in the Western alliance.[8]

Then, is it perhaps the case that here, today, we are experiencing some of the late manifestations of this integration or, at least, one of its echoes? This would suggest that Utrecht, and Europe, are not in fact in *the center*, today. If the center is the American hegemon, then Utrecht is at the periphery of that center, just as it once was a border town on the outskirts of the Roman Empire, with at its central *Domplein* the location of a Roman checkpoint. Then, it is not just the case that historiography seeks to emancipate itself from a Europe centered version of the past; how to write about that past has also increasingly been shaped in places that are not in Europe.

Let us look in more detail both at how the field's center changed from Europe to the US, and at how, in the writing of history, historians struggled to decenter away, so to speak, from their current understanding of science. To do so, I will briefly discuss some of the twentieth century's key figures in the history of science. How do their stories and scholarship illustrate these trends?

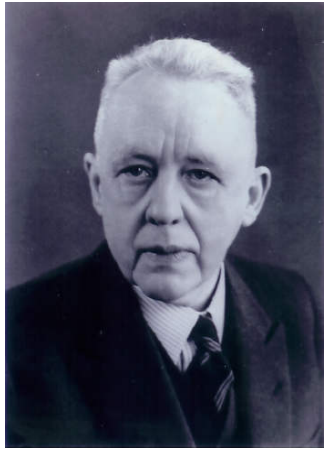

**Fig. 1.** E. J. Dijksterhuis (1892–1965). *Credit*: Museum Boerhaave, Leiden.

Eduard Jan Dijksterhuis (figure 1) was born in 1892.[9] His father was a teacher, and Dijksterhuis, too, would spend most of his professional life teaching mathematics at the public high school in the provincial Dutch town of Tilburg, near the Belgian border. Dijksterhuis had obtained a PhD in mathematics in Groningen. Inspired by the historically tinged classes of famed astronomer Jacob Kaptein, he decided to dedicate himself to the history of science. Despite a long-standing nationalist effort to publish Christiaan Huygens's collected works, this was a barren subject in Holland.

As it was, of course, in most other places. Yet, the ideas of Pierre Duhem appealed to Dijksterhuis—as we learn from his excellent biographer, Klaas van Berkel—and Dijksterhuis's first major publication, *Val en worp*, or *Free Fall and Projectile Motion*, appeared in Dutch in 1924.[10] It had few readers outside Holland. Like Duhem, Dijksterhuis argued that science moved by small incremental steps.[11] Not exactly like Duhem, however, Dijksterhuis insisted on a thorough study of sources, in the original language—not with a goal to unearth the *origins of*, say, Galilean or Newtonian laws of motion in the work of medieval authors, but to understand past scholarship in *its own historical context*. Here, according to the analysis of commentator Floris Cohen, we see a beginning of the attempt to reconstruct "*wie es eigentlich gewesen war*" in the history of science—or, an early attempt to decenter, one could say.

Yet, contemporary notions of what was good science were still central to Dijksterhuis's account, and science's European origins were a *sine qua non* to him. Dijksterhuis believed that there was a clear and universal prescription toward *understanding* of the natural world: science could only begin to offer *explanations* of the phenomena once it had achieved a sufficient level of *mathematization*. This was achieved fully in the Newtonian version of mechanics, which, in turn, proved the pinnacle of *The Mechanization of The World Picture*—the title of Dijksterhuis's widely acclaimed standard work of 1950.[12] In Dijksterhuis's opinion, whatever came after Newton was, essentially, more of the same. He insisted on a thorough study of sources, yes, but at the same time, a general and universal trend toward a particular kind of progress was regarded as a given.

The occurrence of an early modern scientific revolution is actually *not* to be discerned as such in Dijksterhuis's writing: in his perspective, an intellectual continuum connected Newton to the problems and answers of Greek antiquity. Greek antiquity he called "the true nursery of European culture."[13] And in line with his general thesis on advancement by mathematization, innovation by experiment in the sixteenth and seventeenth centuries had been of only secondary importance.

This position resonated with the position he took regarding a secondary school curriculum reform that was hotly debated at the time in the Netherlands. What was the issue? Two types of advanced secondary schools existed: the *Hogere Burgersschool* (the "Advanced School for Burghers," or "HBS") on the one hand, and the classically oriented, more elitist gymnasia on the other. The first offered more sciences and other subjects that were directed at a professional career, whereas the second emphasized schooling in the humanities and classical languages in particular. Only a gymnasium diploma offered direct access to university, and exactly this point was debated throughout the 1920s. Dijksterhuis taught at a HBS, and feared that as a result of these debates its mathematics curriculum would be weakened, just as he feared that those without a steep training in the classics may set foot in the university.[14] He himself had begun his education at his father's HBS, and then had received private tutoring in the classics, before enrolling in mathematics at university.

As a mathematics teacher, Dijksterhuis particularly feared a weakening of a strict deductive style in favor of more intuitive approaches. He further objected to an enlarged role for experimentation in the science curriculum. Both were promoted by progressive school reformers, such as Tatjana Ehrenfest-Affanasjeva, Philip Kohnstamm, and Rommert Casimir. Dijksterhuis abhorred their efforts: only disciplining through encounters with the purest mathematical deductions offered pupils a glimpse of, in his words, the "high moral value" of "strict honesty."[15] At the HBS, the subject of mechanics was taught as a subject of mathematics, and this part of the curriculum stood to be dropped. This would lead to a superficial education, according to Dijksterhuis. Even Galileo, he argued, had depended above all on his mathematical abilities, and not experiment, when clearing up the confusions of Aristotle and his followers. To drop the mathematical treatment of mechanics would mean a return to the misunderstandings of the past. "What all this modernist weakness overlooks," Dijksterhuis thundered, is that "hardly anyone is strong enough to be free.… Sufficiently forceful coercion from above," (that is, "tutelage and exams") was necessary to grow students' self-discipline.[16]

Dijksterhuis may not have been the most light hearted of teachers. Be that as it may, his historiography dovetailed exactly with his ideas about education. So did his larger epistemology. Dijksterhuis was a convinced Platonist and when in 1934 he was asked to draft a national plan for another revamp of the Dutch secondary school system, he handed in a twentieth-century version of the pedagogy from Plato's *Politeia*. Using Euclid's version of geometry and Newtonian mechanics, abstraction and strict logic were to be taught. Experimentation or technical subjects should have no place in advanced secondary schools. Dijksterhuis's plans were driven by Platonist ideals of a steely intellectual elite governing society.[17]

His liberal opponents pointed out that the theory of relativity had just shown both Newtonian mechanics and Euclid to be wrong. Furthermore, it had done so by depending on an empiricist mindset. Dijksterhuis again took his cue from Plato in coming up with an answer: this kind of debate on principles should have no role in education, nor should modern science or experiment interfere with the disciplining virtue of the immersion in strict logical thought. Honesty, reliability, the ability to focus, and clarity of word and mind could only ensue if students had been sufficiently subjugated to fixed ideals and unchallenged truths. Dijksterhuis employed the full force of his historiography to argue against a pedagogy that aimed at intuition, experiment, and visual learning. In his analysis, Euclid had captured Plato's strict idealist epistemology in his geometry. Dijksterhuis, in his own words, "cheered" when, in the early modern period, he saw scholars such as Huygens "conform again to the Greek method. I don't see that as a form of conservatism," Dijksterhuis elaborated, "but as the realization of the eternal insights that underlie Greek rigour."[18]

Just as Plato imagined that an intellectual elite should govern, Dijksterhuis had few democratic instincts. Even if, according to his biographer van Berkel, he was not very active politically, and could only scold at the tastelessness of German Nazis and their Dutch equivalents, and objected to anti-Semitism, Dijksterhuis did briefly join a marginal fascist group at the beginning of the Nazi occupation of the Netherlands. This was the National Front, led by a local man from the Tilburg area, Arnold Meijer. When this group revealed its anti-Semitic instincts, Dijksterhuis quickly terminated his membership. However, he was not too unsettled by the German occupation; he just regretted the spiraling dynamic of repression and resistance. Unaware of the gravest acts of violence, it was not so much that Dijksterhuis disagreed with the "new ordering of Europe"; he rather thought it was led by the entirely wrong kind of people.

In 1944 he accepted a position at the University of Amsterdam. This was the first salaried university position in the history of science in the Netherlands. Accepting such a position was controversial since due to the occupation many faculty jobs had been vacated, sometimes forcibly. By this time, the only students who could follow courses had signed loyalty oaths to the occupying forces—others were often living in hiding for fear of being deported to Germany as forced laborers, or worse.

Dijksterhuis held that teaching simply needed to continue and that therefore accepting a position was an apolitical act.[19] Yet only a year later in 1945, after the liberation of the Netherlands by the Allied forces, the reinstated Dutch officialdom judged this very differently. An official committee condemned Dijksterhuis for his "grave error" and his position in Amsterdam was terminated.[20] Dijksterhuis returned to teaching high-school mathematics in Tilburg. And, eventually, within a few years, the national mathematics curriculum would be reformed exactly opposite to his Platonist ideals. His lectures intended for Amsterdam's students were rewritten into the book mentioned earlier, *The Mechanization of the World Picture*. Upon its appearance in 1950, it immediately attained the status of a classic text: it was translated in a multitude of languages—published in English by Oxford University Press and in German by Springer.[21] The book was awarded the P. C. Hooft-prize, the most prestigious award for any Dutch literary work. This initiated Dijksterhuis's rehabilitation, and he was soon offered a professor's chair in history of science in Utrecht, again a first, and a position that counts as the beginning of Utrecht's tradition in the field.[22]

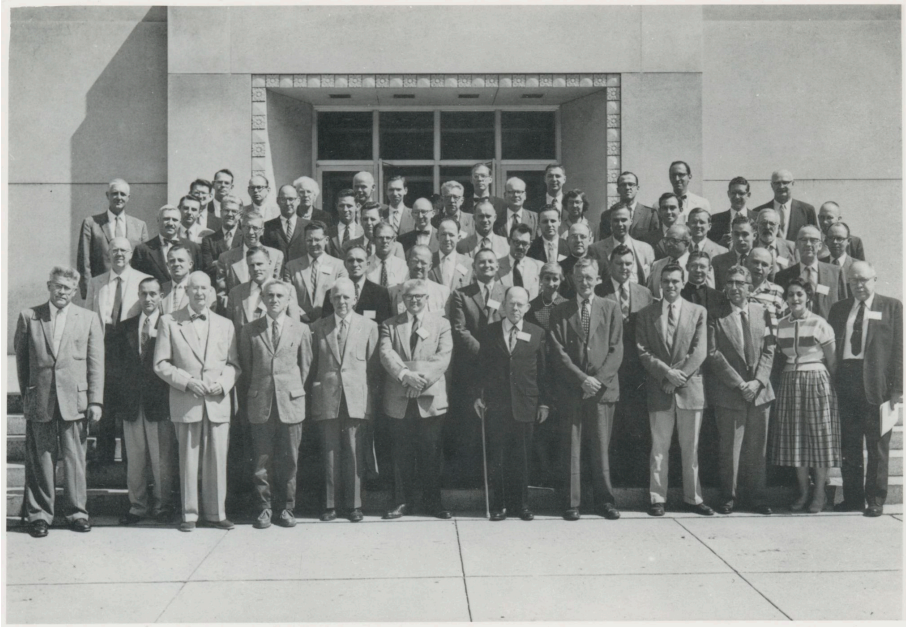

**Fig. 2.** Participants at the conference "Critical problems in the history of science" (Madison WI), September 1–11, 1957. Dijksterhuis is fifth from the left in the front row; Thomas Kuhn is third from the right in the top row.

Dijksterhuis was awarded the HSS Sarton medal in 1962. He had traveled to the United States for the first and only time a little earlier, in 1957, when he attended the highly influential conference *Critical Problems in the History of Science* in Madison, Wisconsin (figure 2).[23] He was positively impressed by the level of scholarship presented there. Yet, there was also enough to complain about: such as speakers who, "monotonously read out a large number of sheets," while practicing an "abusive" mispronunciation of foreign language names (such as "Layvoshay" for Lavoisier) and choosing Troia's Steak House as the destination for breakfast, lunch, and dinner, every single day.[24] At the conference itself, the historiographical concept of an early modern scientific revolution, which had been prominently brought into circulation by Herbert Butterfield and Alexandre Koyré, was heavily debated.

"In history of science," Dijksterhuis observed, "the United States is now, without a doubt, first in the world." It had most professorial chairs, and these were filled with quality people. This expansion of the field had been fueled by Cold War funding levels for the sciences, and the accompanying wish to promote a pro-science public culture. At Madison, Dijksterhuis had seen many young scholars with sharp minds—Stillman Drake and Thomas Kuhn (figure 3) particularly stood out to him. He enjoyed sharp discussions until late at night with Kuhn, and was convinced that this was, and I quote, a "young and many sided historian of science of whom we most certainly will still hear much."[25] Dijksterhuis returned to Europe, while someone like Koyré would stay in the United States, as George Sarton had before him. If their generation was the first in professional history of science, the second would be dominated by Americans; most prominently, of course, by Kuhn and his book *The Structure of Scientific Revolutions*.[26] The professional center thus was no longer "in Europe"; it had moved to the United States, even if the subject itself was still squarely "European."

How did Dijksterhuis imagine the public role of the history of science? He was convinced that the strong separation between the sciences and the humanities, particularly as this was engrained in the Dutch secondary school system, was a grave social ill. The study and teaching of the history of science was essential to amend it. Yet, debates in which Dijksterhuis participated about how mathematics and the sciences should be taught also brought out another, less explicit element of his vision for the history of science.[27] To Dijksterhuis, those debates needed to be informed by historical argument: it mattered greatly, for instance, what the role of experiment had been in Galileo's science when designing a school curriculum. Finally, the values that Dijksterhuis' ideal school curriculum emphasized reflected a social ordering that was grounded in his Platonist views. These, in turn, were justified by the historical course of mathematization and mechanization that the progress of the sciences revealed—thus, historiography of science could justify political ideology. For Dijksterhuis, it justified the idea of governance by an intellectual elite.

Such a connection between historiography and preferred social ordering was hardly unique to Dijksterhuis, his era, or his conservative position; one can develop a similar analysis for Marxist astronomer and historian Anton Pannekoek, or, in fact, for Manhattan Project veteran James Bryant Conant, who emphasized the free exchange of ideas in his attempts at historiography, while promoting the teaching of the history of science to strengthen free and open democracy.[28] Conant famously hired physics graduate Thomas Kuhn as a teaching assistant for his course at Harvard in the late 1940s, thus drawing the latter into the subject. A decade later, Kuhn traveled to Madison, Wisconsin, to debate Dijksterhuis.[29] It is to Kuhn we turn to now.

Kuhn's philosophy of science is well known: in paradigmatic normal science, certain anomalies occur that may lead to crisis, revolution, a new incommensurable paradigm, and normal science again. Kuhn was at least as much a child of his time as Dijksterhuis; for example, he had been influenced by the child development theories of Jean Piaget, ideas about the early modern scientific revolution à la Koyré, and the communist capturing of minds feared by American Cold War culture warriors.[30]

Whether the seventeenth-century scientific revolution was a Kuhnian revolution can be debated, and it very much has been; but it is clear that to Kuhn, revolutionary things happened in the early modern study of nature.[31] Kuhn himself mentioned a kind of epiphany he had experienced when assisting Conant in teaching history of science: reading Aristotle, he shockingly discovered that his own Newtonian expectations were blocking him from seeing the consistency and integrity of Aristotle's physics. This experience put him on the path to his famous book, and its concept of paradigms. As he wrote to Conant in 1961: "Newton was not trying to do Aristotle's job better; rather, Aristotle had been trying to do a different job and one that Newton did not do so well."[32] In other words, Aristotle and Newton had been working in completely different, incommensurable paradigms.

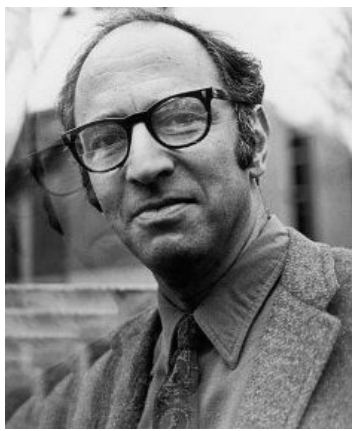

**Fig. 3.** Thomas Kuhn (1922–1996). Photo by Bill Pierce, *Credit*: Wikimedia Commons.

A point of this kind might well be leveled against much of what Dijksterhuis had argued. A belief in the cumulative nature of science had stood in Dijksterhuis' way, one could argue. As Kuhn wrote to Conant, the example of Aristotle and Newton illustrated that "science was not cumulative."[33] The belief in the cumulative nature of science elevated the present—for Dijksterhuis, that was science à la Newton—to a preferred vantage point that obscured true historical insight. In other words: Kuhn held that decentering from the now, the position we have arrived at, is necessary, and doing so brought one to recognize paradigms and revolutions. To achieve this decentering, Kuhn emphasized that a thorough study of sources was necessary, and this emphasis was one that he did share with Dijksterhuis.

Let us consider one such example and investigate how Kuhn argued for a de-centering of the now, while paying appropriate attention to the sources. This is a different example than debates about the seventeenth century. The example I would like to discuss is the contentious case of Max Planck's black body radiation formula and how Planck interpreted it.

In 1900 Max Planck (figure 4) had introduced the following expression for the distribution of energy over radiation frequencies, for a given temperature:

$$\rho(\nu, T) = \frac{8\pi h \nu^3}{c^3} \frac{1}{e^{h\nu/kT} - 1}$$

This formula captures how the energy density $\rho$ of electromagnetic radiation in a cavity is distributed over frequencies $\nu$ at temperatures $T$ for radiation of a perfect emitter or absorber, in other words, for a perfect black body. All the other symbols in the expression are constants, including the novel constant *h*. This relation had been much sought after, since it captured what was universal in the relation between the absorption and emission properties of any physical body. Observations had suggested a particular kind of graph (figure 7), but it was not clear what formula might fit that graph until Planck famously made the above proposal. That formula has commonly been considered the starting point of quantum physics.

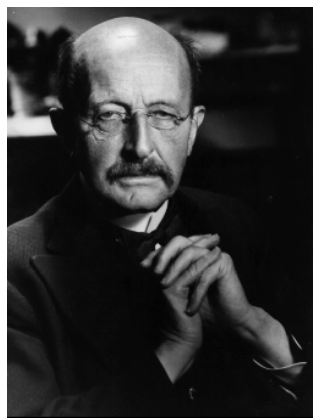

**Fig. 4.** Max Planck (1858–1947) in his later years. *Source*: Max-Planck-Gesellschaft.

The formula entailed—as all agree that Albert Einstein made unequivocally clear in 1905 and 1906—that the radiation energy was confined to discrete elements of size *hv*, instead of continuous elements that could take on any value. According to Kuhn, however, this was actually not how Planck himself interpreted his formula. Planck's notes were lost in World War I, so a detailed reconstruction using a wealth of sources is impossible, but by piecing together various statements in a number of Planck's publications, Kuhn, in the early 1970s, settled on the point of view that Planck had originally believed that, in fact, the energies in the radiation system were *not* really quantized, *not* really discretely valued. Despite how we understand Planck's formula today, quantization had only been a convenient assumption to aid calculations. In fact, according to Kuhn, Planck in 1900 believed that the radiation energies were to be found somewhere inside energy boxes of size *hv, 2hv, 3hv*, et cetera. In Kuhn's novel reading, Planck did not think that the energies were in actual fact confined to the discrete *boundaries of these boxes* at the discrete values of *hv, 2hv, 3hv*, etc. Strictly speaking, this interpretation of the radiation formula was too liberal when compared to the validity criteria of the assumptions of the theory from which it was inferred, the statistical theory of atoms of Ludwig Boltzmann, its nineteenth century master. Yet, the invalidity of this interpretation of Planck's own formula would only become clear if one were to consider what happens to it at high radiation energies versus temperatures—and this, Kuhn pointed out, was only made explicit by Einstein much later, in 1910.[34]

Planck's would then be a conservative position, since the continuities of nineteenth century physics would then still have survived in his mind, despite his own formula. This kind of conservatism resonates well with the image that one has of Planck. He was the son of a Professor of Law, the grandson of a Professor of Theology, and the signatory to a defamed and prominent nationalist manifesto during World War I. He soon became a prominent leader and administrator of German science, in many capacities. Planck was an establishment figure, not someone who sought to rock the boat.[35] Kuhn's version of Planck's physics of 1900 seemed to fit the man, at least as he came to be known and remembered.

Thomas Kuhn wrote up his ideas in a book that appeared with Oxford University Press in 1978.[36] Surprisingly, it largely avoided the vocabulary of paradigms and anomalies. It may still be taken as a typically Kuhnian account, however: there is a certain consistency in Planck's thought, a consistency that fits nicely with the notion that Planck was taking his cues from a full and theoretical paradigm—the paradigm of nineteenth-century physics with its continuous, and not discrete radiation fields.

Before the publication of Kuhn's book, the uncontested historical account of Planck had been given by Martin J. Klein (figure 6), a contemporary of Kuhn.[37] Klein had also been educated as a physicist. He had graduated from MIT in 1948 and had continued as a faculty member in physics at the Case Institute in Ohio. Klein traveled to Europe, in fact to Holland, where he spent the 1958–59 academic year as a Guggenheim fellow at the University of Leiden.[38] This offered him the opportunity to further develop his interest in the history of atomic theories for thermodynamics. Paul Ehrenfest, a student of Boltzmann, had taught the subject at Leiden, and Klein befriended his widow, Tatjana-Ehrenfest Afanasjewa—whom we already encountered as one of Dijksterhuis's liberal opponents in debates about how to teach high-school mathematics. Marty, as Martin Klein was mostly known, first edited Paul Ehrenfest's collected papers, and then set out to write his biography.[39] Along the way, he contributed key publications on Planck, Einstein, and other physicists, particularly on the early years of the quantum theory, as it was being shaped at the hands of these authors, often under the Socratic criticism of Ehrenfest. Through this work, Klein turned into an historian. He would become General Editor of the *Collected Papers of Albert Einstein*, and, in 1967, Eugene Higgins Professor in the History of Science at Yale University; in 2005, Klein was the first recipient of the Abraham Pais Prize for the History of Physics, awarded by the American Physical Society. He remained a regular visitor to Holland, eventually exchanging Leiden for Amsterdam as destination.

I met Klein in Amsterdam as a student in 1999 and we would have many more occasions to interact. Generous and kind, he invited me to stay at his house for a week in 2006, when I helped him organize some of his papers—and then, going through one of his boxes, I came across a photograph that, in some subliminal way, must have led to tonight's lecture: it had Dijksterhuis seated on some throne, jokingly flanked on either side by a kneeling Thomas Kuhn and Martin Klein; all having a good time. I have tried frantically to unearth that photo in the last couple of months, but unfortunately have not succeeded. Martin could no longer help me. In 2009, ten years ago, Martin Klein passed away.

Let us return to Planck. In Klein's account from the early 1960s, Planck was brought—forced—to accept the discrete energy quantum, because it gave the desired formula. In Klein's version, Planck did not study taking a continuum limit, overlooking this step in the work of

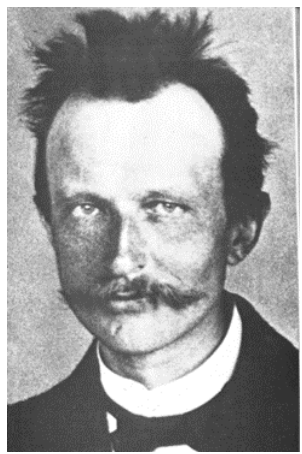

**Fig. 5.** Max Planck (1858–1947) in 1901. *Credit*: Max-Planck-Gesellschaft.

Boltzmann, which prevented him from seeing how his formula would be connected to nineteenth-century physics and making him stick with novelty—in particular as this captured observations.

According to Klein, discontinuity and revolution would have been forced upon Planck by his own formula, whereas according to Kuhn, Planck retained the old nineteenth century continuum view of nature, yet unaware of the problems this may pose, even for his own formula. In 1906, six years after Planck had proposed his formula, he did express the continuum point of view[40]—adding consistency to Kuhn's account, and a little creative inconsistency to Klein's Planck. Kuhn's Planck was the classical conservative, fitting Planck's larger image. Klein's Planck was not so conservative—radical, with a hint of crazy, perhaps better fitting his formula of 1900, than the persona as he later became known. Yet, did that image also capture Planck in 1900 (figure 5)?

When Kuhn's book on Planck appeared in 1978, he and Klein had been friends for a good twenty years. How could they not be? They were both American Jews from urban professional families, born only two years apart and with vibrant personalities, coming up as junior physicists in the context of World War II and finally arriving as some of the world's most prominent historians of science and nearby Ivy League colleagues. Yet, their differing reconstructions of Planck put a heavy strain on their relationship, affecting it, ultimately, beyond repair.

Kuhn began sharing his alternative reconstruction with Klein early in 1972: "I have now reread more Klein and thought through my own plans," Kuhn wrote to Klein in March of that year. "I think I am on the track of a significant new version of the 1900–1911 story. What I suspect can be made plausible is this: contrary to prevalent impression, Planck never thought of restricting oscillator energies to *nhv* until 1910. I am not sure if this view is going to hold up, but I am going to give it a run."[41] In the same letter, Kuhn urged Klein to republish his work on Planck that had been scattered across a number of venues, some of which by then had become less visible: "Individually and collectively," Kuhn stated, "your Planck papers are the best there is, and they are also damned good history of science by any standards." Klein, however, did not follow up the suggestion.

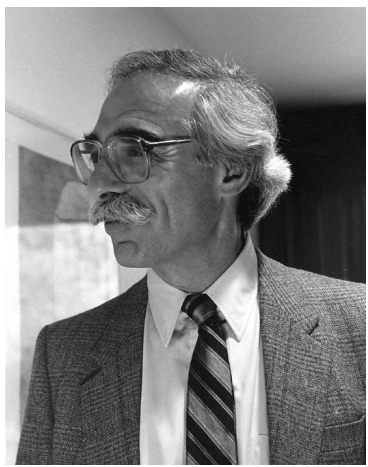

**Fig. 6.** Martin J. Klein (1924–2009). *Credit*: National Academy of Sciences.

He also declined to read a draft of Kuhn's book two years later. He explained to Kuhn why: "I have felt somewhat defensive in some of our discussions during the last two years, and it seems to me that you have been that way also. This mutual defensiveness is not good for our friendship or for us individually and I think my reading and commenting on your manuscript might well aggravate matters. In any case, I have been disturbed by how our discussions have been going and am glad finally to have said so."[42] Kuhn was grateful for this letter, and accepted Klein's decision "fully and unequivocally." Yet, he also suggested the possibility of finding, as he wrote to Klein, "some other mode of interaction. My primary reason for hoping you would read the manuscript was that our going over it together *might* have provided a means of facing our joint problems and eliminating or relaxing them. That was not the only possible outcome and perhaps not even the most likely. I don't want to urge it on you again, but neither am I entirely happy about postponing the problems during the time it will take for my book to appear."[43]

That, however, is exactly what happened. And another challenge to their friendship was added in 1974. Tom Kuhn had lectured in Japan on Einstein's ideas on specific heats and their relation to the early quantum theory. He needed to be reminded by Klein that he had essentially just rehashed some of his, that is, Martin Klein's work, without giving it any reference.[44] Kuhn actually agreed, and was greatly embarrassed: "I have slipped badly," he acknowledged to Klein, "and it should not have occurred. I … am much upset." Kuhn asked his Japanese editor for a note to be added to his manuscript that fully acknowledged the overlap with Klein, and suggested that he ought not to have presented his paper, given that overlap.[45] The book on Planck was still to appear.

Which it did in 1978. *Black Body Theory and the Quantum Discontinuity* was reviewed by Martin Klein for *Isis*. This review was combined in a "Symposium" section with reviews of the same book by Abner Shimony and Trevor Pinch.[46] Klein's review was not positive. Thomas Kuhn thought it "unfair and ungenerous," if not "extraordinarily onesided," as he wrote to Klein.[47] Kuhn thought Klein had misrepresented the sources that substantiated his reading of Planck. Thus, Klein would have ended up having Planck do an awkward U-turn between 1900 and 1906, because in 1906 Planck had of course lectured in line with what soon came to be described as classical physics. "Your … account of Planck's development makes no sense," Kuhn insisted.[48]

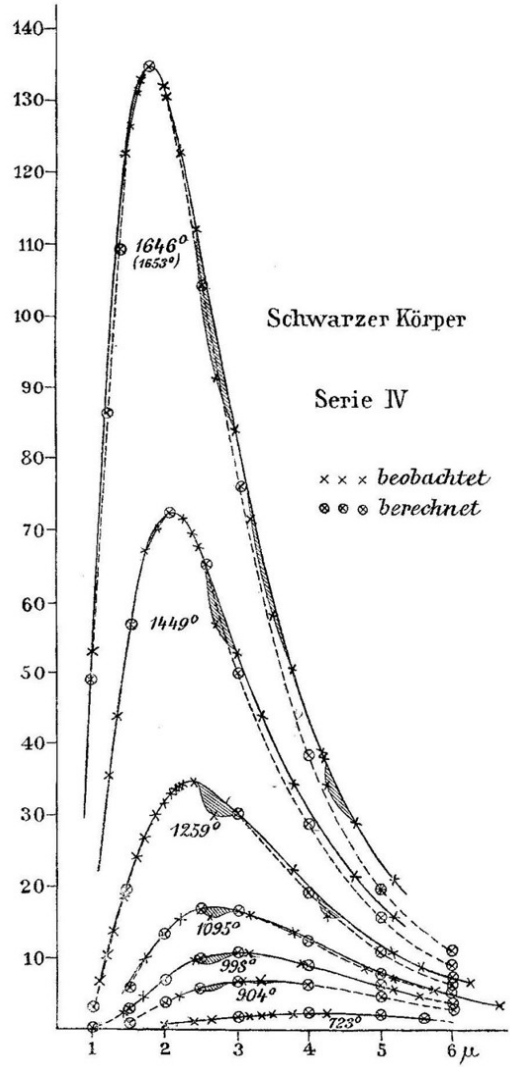

**Fig. 7.** Black body radiation at various temperatures as measured by E. Lummer and E. Pringsheim, with deviations from the expected Wien law indicated; Planck offered a new law that would capture all data. Source: E. Lummer and E. Pringsheim, "1. Die Vertheilung der Energie im Spectrum des schwarzen Körpers und des blanken Platins; 2. Temperaturbestimmung fester glühender Körper," *Verhandlungen der Deutschen Physikalischen Gesellschaft* **1** (1899), 215–35, on 217.

Klein, in reply as well as in the review, pointed out that Kuhn had yet again failed to properly acknowledge some of his earlier work. "I find that unfair and ungenerous of you," Klein added in a letter.[49] Kuhn yet again conceded this point, but now disqualified it as "thorny underbrush."[50] Klein further reminded him that in those overlooked publications, he had argued how Planck had studied more Boltzmann between 1900 and 1906, and that with this added knowledge, one could still very well make the point that Planck wavered between discrete and continuous views of the radiation process in those years, just as he, Klein, had in fact argued. Finally, in his book Kuhn had stated that Planck's formulas and sentences "need not be read literally." Klein thought that this was entirely unsatisfactory, and he judged Kuhn's treatment of Einstein, Ehrenfest, and others in similar terms.[51]

Kuhn did not just disagree with Klein; he felt that an entirely different approach to the history of science divided them. "Our basic difference seems to me historiographic and quite deep," he wrote to Klein. "In conversations with me, you have repeatedly referred to both Planck and Boltzmann as 'confused.'" Kuhn argued that such "confusion" was not to be attributed to historical actors in the way that Klein did. Kuhn elaborated by referring to "the main maxim" he used when teaching his "beginning students: in approaching a significant historical figure, look first for passages that don't make sense; then try to find a way of reading that makes plausible that a person of great intelligence would have written them; finally, look back and notice how the significance of previous unproblematic passages has changed." Kuhn concluded: "confusion is inevitably reduced as a result."[52] Reading between the lines, one might say that Kuhn was arguing that fully reconstructing the paradigm from which a historical actor operated clarifies his thought.

According to Kuhn, Klein's reference to confusion was to be found "in the relation between what the texts appear literally to say and the interpretation of those texts provided by the historian." By working this way, Kuhn wrote to Klein, "allowance [is] made for contradictions *within* the thought patterns reconstructed by the historian. I … believe you attribute to early papers doctrines that did not come into existence until later.… Yes, the relation between their texts and the views *you attribute to* Boltzmann and Planck is confused. But the existence of those confusions and contradictions does not show that Boltzmann or Planck were confused. For that … one must show that Planck and Boltzmann actually held the views attributed to them." Kuhn added that the attribution of later insight to historical actors follows from a "sleepwalker view of scientific development," and may happen, as in Kuhn's opinion it did in Klein's work, when in studying the sources insufficient attention is paid to detail.[53] One might say that Klein had insufficiently "decentered" the modern view of quantum theory in his account of Planck—at least in Kuhn's opinion.

Predictably, Klein took exception: "for some reason you just don't see what is actually in my work," he contended. "You find it appropriate … to read me a little lecture about how one should do history, reminding me of what you tell your beginning graduate students and suggesting that I do not pay enough attention to the sources themselves. I don't know why you should take that tone with me. I find it hard to avoid the conclusion that you do not really take my work seriously, despite all your statements to the contrary."[54]

Looking at this episode, from the awkward invitation to republish his earlier papers on Planck through the back and forth on the book and its review, one senses that Martin Klein refused to be Kuhn's straw man. And rightly so: even if Kuhn's account of Planck is today considered most convincing by the majority of specialists, debates about its differences with Klein's have preoccupied a fair number of some of our best colleagues, and indeed have done so for a good number of years.[55]

In his letters to Klein, Kuhn did articulate a view that has become the essential perspective in historiography of science. That is the requirement to decenter; to sufficiently question one's perspective and recognize its conditioning in the here and now; to recognize how such conditioning clouds judgment and understanding of the past and how the present has materialized from it. Yet, Klein's work was ill cast for the role of antithesis to this kind of scholarship. Klein sensed he was being put in that role and resisted it. Even if he was critical at times of Kuhn's historiography by paradigms and revolutions (Kuhn even placed the true source of their tensions there), Klein undoubtedly acknowledged a larger message to decenter and, foremost, to adhere as closely as possible to historical sources—*No Truth, Except in the Details* is the title of the edited volume that honored Klein's career achievements.[56] Indeed, in his reply to Kuhn, Klein did just that: he appealed to the authority of the sources—to his reading of them at least. He did not engage in a discussion of a larger historiographic perspective.

The acknowledgment that to decenter was the larger direction that the history of science had taken and the path it would continue upon is reflected in the standard works that came after Kuhn in the 1980s—also when they were critical of Kuhn's point of view. Kuhn's historiographical criticism would have been justifiably directed at Dijksterhuis and others of that generation. It is also at these authors that *The Structure of Scientific Revolutions* originally took aim. But this charge leveled at Klein left the latter in an awkward position. In the end, he and Kuhn literally decided to "agree to disagree,"[57] and again found a working relation. Klein, however, despite his general editorship of Einstein's collected papers, eventually hit upon a writer's block that effectively obstructed him from finishing the second volume of his celebrated Ehrenfest biography. When I, as a young postdoc, spent that week in his home, going through old photos, notes, and drafted chapters, and we hit upon the image of himself and Thomas Kuhn, flanking Dijksterhuis seated on his would-be throne in the late 1950s, Marty still did not feel like returning to his debates on Planck—and I did not insist, as the recollection obviously pained him.

So, ever again unseating the perspective of the here and now, decentering the self, has been a recurring red thread of innovation in the history of science, at least since the helm passed from the generation of Dijksterhuis to that of Kuhn and Klein. Where does that leave us now, and here? How might that perspective aid the public? Does it actually imply a perspective that seeks to undermine the authority of the sciences, as has often been suggested? And what may it tell us about being "in Europe"?

Let us again address the public role of the history of science. The notion that science studies have been unfairly critical of the sciences is entirely maladroit: that notion itself in fact derives from a Whig notion of the importance of the present state of science. It denies science's contingent character. At the same time, we must observe that being able to situate knowledge in its proper time and place is hugely important at this very moment, for this is the time of fake news and alternative facts. Understanding how scientific evidence and debate work is of vital importance to the public at large, in Europe at least as much as anywhere else—one need only think of the controversies about vaccination and climate science as two examples that are in the papers daily. And how could they not be: we just experienced the hottest summer in the Netherlands on record.

Historians of science must engage the public and offer their insight into how debate and consensus work in science, because scientists themselves and their communications officers are not getting the job done—their Whiggishness may be getting in the way. Fortunately, many of us are doing just that: think for instance of the wonderful book, *Merchants of Doubt* by Naomi Oreskes and Erik Conway on tobacco science and how it relates to climate debates.[58] I also wish to mention work by a young colleague of mine from Amsterdam, Sjang ten Hagen, who has studied the rise of the notion of "fact" in nineteenth century German contexts:[59] it is this kind of scholarship that reveals the situatedness of knowledge on issues that matter today. But we need to engage more and more directly, not just by writing books but by showing up in the raw public fora of debate.

Where does today's history of science leave us, here, in Europe? Perhaps as subject less central and as source of scholarship less dominant. Is this a reason for worry, frustration, or malcontent? I say no: the circumstance that European scholarship has been less dominant is an inevitable consequence exactly of the vital public role of history of science: societies across the globe recognize its worth, in either teaching, scholarship, or debate. Besides, we have seen here this week much excellent European scholarship and the diminishing role of European scholarship has always been a relative and never an absolute effect—one may even begin to think in terms of its reversal, particularly since the creation of the Max Planck Institute. In any case, it is exactly because history of science has an important role on the global stage that we see people from all over the world at such events as the HSS conferences. I applaud the fact that the HSS recognizes it should also serve those beyond North America, and we should celebrate its presence here. Besides, with so very many excellent contributions this week, and such wonderful institutions as, indeed, Utrecht University, we hardly have reason to complain at all, in Europe. Finally, it is a good thing, I say, that historians are "provincializing Europe" in their effort at decentering the here and now: it derives from a vital question—why and how did European reason come to be seen as self-evident? That question reflects and justly serves a long overdue rise of global pluralism, and that pluralism I welcome.

**Acknowledgments:** I am grateful for comments by Don Howard, Ted Jacobson, Michel Janssen, Joe Martin, Dennis Lehmkuhl, Jürgen Renn, and Richard Staley. I further thank David Baneke and Ariane den Daas of Utrecht's Descartes Centre; Jay Malone and Bernard Lightman of the HSS, and the Elizabeth Paris Endowment for Socially Engaged History and Philosophy of Science.

[42] Martin J. Klein, letter to Thomas S. Kuhn, October 7, 1974, MJK.

[43] T. Thomas S. Kuhn, letter to Martin J. Klein, July 28, 1974, MJK.

[44] Martin J. Klein, letter to Thomas S. Kuhn, July 29, 1974, MJK; Klein's article is: Martin J. Klein, "Einstein, Specific heats, and the Early Quantum Theory," *Science* **148**, no. 3667 (1965), 173–80.

[45] Thomas S. Kuhn, statement dated July 31, 1974, shared with Martin J. Klein in letter to the latter, August 1, 1974; Thomas S. Kuhn, letter to Tetu Hirosige, August 7, 1974, MJK. Both Klein and Kuhn presented articles at the 14th International Congress of the History of Science, held in Tokyo and Kyoto, August 19–27, 1974.

[46] Martin J. Klein, Abern Shimony, and Trevor J. Pinch, "Paradigm Lost?: A Review Symposium," *Isis* **70**, no. 3 (1979), 429–40; Klein's review is found on 430–34.

[47] Thomas S. Kuhn, letter to Martin J. Klein, August 1, 1979, MJK.

[48] Thomas S. Kuhn, letter to Martin J. Klein, August 1, 1979, MJK. Some of the points raised in the exchange with Klein are included in Kuhn, "Revisiting Planck" (ref. 34).. Here, Klein likely features as Kuhn's "most learned and authoritative critic" (p. 237), who would have conceded that Planck argued in line with a continuum theory in 1906.

[49] Martin J. Klein, letter to Thomas S. Kuhn, September 4, 1979, MJK.

[50] Kuhn, cited in Thomas S. Kuhn, letter to Martin J. Klein, August 1, 1979; for conceding this point, see Kuhn's letter to Klein of October 18, 1979; MJK.

[51] Martin J. Klein, letter to Thomas S. Kuhn, September 4, 1979, MJK.

[52] Thomas S. Kuhn, letter to Martin J. Klein, August 1, 1979, MJK.

[53] Thomas S. Kuhn, letter to Martin J. Klein, August 1, 1979, MJK.

[54] Martin J. Klein, letter to Thomas S. Kuhn, September 4, 1979, MJK.

[55] The most recent authoritative review is Anthony Duncan and Michel Janssen, "Planck, the Second Law of Thermodynamics, and Black Body Radiation," in *Constructing Quantum Mechanics*, vol. 1, *The Scaffold: 1900–1923*, ch. 2 (Oxford: Oxford University Press, 2019). For earlier contributions, see Allan A. Needell, *Irreversibility and the Failure of Classical Dynamics: Max Planck's Work on the Quantum Theory, 1900–1915* (PhD dissertation, Yale University, 1980); Peter Galison; "Kuhn and the Quantum Controversy," *British Journal for the Philosophy of Science* **32**, no. 1 (1981), 71–85; Olivier Darrigol, *From c-Numbers to q-Numbers: The Classical Analogy in the History of Quantum Theory* (Berkeley: University of California Press, 1992); Jochen Büttner, Jürgen Renn and Matthias Schemmel, "Exploring the Limits of Classical Physics: Planck, Einstein, and the Structure of a Scientific Revolution," *Studies in History and Philosophy of Modern Physics* **34**, no. 1 (2001), 35–59; Clayton Gearhart, "Planck, the Quantum, and the Historians," *Physics in Perspective* **4**, no. 2 (2002), 170–215; Massimiliano Badino, *The Bumpy Road: Max Planck from Radiation Theory to the Quantum (1896–1906)* (New York: Springer, 2015).